\documentclass[12pt]{article}
\usepackage{graphicx}
\usepackage{amsmath}
\usepackage{amssymb}
\usepackage{amsfonts}

\newcommand{\beq}{\begin{equation}}
\newcommand{\eeq}{\end{equation}}
\newcommand{\bea}{\begin{eqnarray}}
\newcommand{\eea}{\end{eqnarray}}

\newcommand{\vepsi}{\epsilon}

\newcommand{\hhat}{\hat{h}}
\newcommand{\bhat}{\hat{b}}

\newcommand{\zhat}{\hat{z}}
\newcommand{\thetahat}{\hat{\theta}}

\newcommand{\para}{\parallel}
\newcommand{\half}{\frac{1}{2}}
\newcommand{\wt}[1]{\widetilde{#1}}
\newcommand{\mbf}[1]{\mathbf{#1}}

\newcommand{\del}{\nabla}
\newcommand{\divr}{\nabla \cdot}
\newcommand{\curl}{\nabla \times}
\newcommand{\DDt}[1]{\frac{D\, {#1}}{D t}}
\newcommand{\ddt}[1]{\frac{\partial\, {#1}}{\partial t}}
\newcommand{\ddr}[1]{\frac{\partial\, {#1}}{\partial r}}

\newcommand{\oover}[1]{\frac{1}{#1}}
\newcommand{\dext}{\mathrm{d}}
\newcommand{\ptild}{\widetilde{p}}

\newcommand{\MoH}{\frac{M}{H}}
\newcommand{\partialt}[1]{\partial {#1} / \partial t}
\newcommand{\partlxy}[2]{\partial {#1} / \partial {#2}}
\makeatletter
\def\Tbar{\mathchoice
   {\TTbar\displaystyle\textstyle{-}}%
   {\TTbar\textstyle\scriptstyle{-}}%
   {\TTbar\scriptstyle\scriptscriptstyle{-}}%
   {\TTbar\scriptscriptstyle\scriptscriptstyle{-}}%
   \!T}
\def\TTbar#1#2#3{{\setbox0=\hbox{$#1{#2#3}{\mathrm{T}}$}
     \raise2\p@\vbox{\hbox{$#2#3$}}\kern-.35\wd0}}
\makeatother

\begin{document}


\title{Non-stationary magnetized axially symmetric equilibrium from the fluid equations of motion}
\author{Robert W. Johnson\footnote{Alphawave Research, Atlanta, Georgia, USA. (robjohnson@alphawaveresearch.com)}}

\date{\today}

\maketitle

\begin{abstract}
The equations of motion for a fully ionized hydrogenic plasma in applied coaxial electric and magnetic fields are analyzed, where the term for the Hall effect in the generalized Ohm's law equation picks up a factor of 1/2 relative to its usual expression.  Magnetization of the medium is incorporated through the decomposition of the Hall term and the inclusion of the magnetization force, which is found to equal or exceed the gradient of the scalar pressure.  A limit on the kinetic pressure obtains which corresponds to the usual limit of unity for a certain selection of parameters.  Solutions of these equations for the free motion of the charges in the case of an infinite column with azimuthal symmetry are compared for various prescribed pressure profiles, where one finds that the profile near the outer edge plays an important role in the feasibility of the equilibrium.
\end{abstract}

\newpage

\section{Introduction}
%

The analysis of the many-body system commonly called a plasma is the marriage of electrodynamics and fluid mechanics.  In the classical regime, one considers Maxwell's theory of electromagnetic fields~[\cite{maxwell-1864}] and Newton's theory of dynamics~[\cite{newton-princ}] as expressed by the moments of the Vlasov equation~[\cite{vlasov-68,dendybook-93,staceybook05}].  For this article, we consider the application of electromagnetic hydrodynamics (EM-HD) to a fully ionized medium of a single hydrogenic species in the neutral fluid limit, with consideration of the convective terms and inclusion of the magnetization force.  The distinction between the free and bound current is maintained formally by isolation of the fluid and drift velocities.  Collisional disruption of the gyromotion is modelled by a simple correction factor applied to the species magnetic moment and gyro-momentum.  The pressure across an infinite, symmetric column of unit radius is prescribed, and the equations of motion are solved for the motion of the charges as represented by the free and bound momentum and current densities.  The solutions for the free momentum are sensitive to the pressure profile near the outer edge, and a limit on the kinetic pressure obtains from the nonlinear magnetization model.

Here we consider the fluid description of a neutral, $z_i = 1$ plasma of species $s\in\{e,i\}$ with total particle density $n \equiv \sum_s n_s = 2 n_0$, mass density $\rho_m \equiv \sum_s n_s m_s = n_0 (m_e + m_i)$, charge density $\rho_e \equiv \sum_s n_s e_s = 0$, free momentum density $\rho_m \mbf{V}_f \equiv \sum_s n_s m_s \mbf{V}_s = n_0 (m_e \mbf{V}_e + m_i \mbf{V}_i)$, free current density $\mbf{J}_f \equiv \sum_s n_s e_s \mbf{V}_s = n_0 e (\mbf{V}_i - \mbf{V}_e)$, and pressure $p \equiv n\, \Tbar \equiv \sum_s n_s \Tbar_s \equiv \sum_s p_s = n_0 (\Tbar_e + \Tbar_i)$ for $\Tbar \equiv k_b T$.  Note that the plasma density $n_0$ is 1/2 the total particle density $n$, that energy equipartition gives $W_s^\perp + W_s^\para = 3 \Tbar_s / 2$ such that $\mbf{v}_s^\perp = \sqrt{2 \Tbar_s / m_s}$, and that $\mbf{V}_{e,i}$ are the species fluid or guiding center velocities.  Addressing in turn the equations for the electromagnetic fields, mass, energy, and momentum, we arrive at a model which predicts the equilibrium current and plasma flow from the prescribed pressure and applied electric and magnetic fields.

\section{Theoretical model}\label{sec:theomodel}

\subsection{Maxwell's field equations}

The classical theory of electromagnetism has been well established, and suitably quantized~[\cite{halzenmartin,ryder-qft}], its predictions have been experimentally verified to the level of parts per billion~[\cite{2006PhRvL..97c0801O,Hagiwara:2006jt}].  The energy density of a magnetized plasma falls between that of everyday experience and that of high-energy collisions, and there is no reason to believe {\it a priori} that the theory should fail for a fully ionized medium.  Formally, one may express the equations \bea \label{eqn-Maxwell}
\divr \mbf{E} = \rho_e / \vepsi_0 \;,& &\curl \mbf{B} = \mu_0 \mbf{J} + \mu_0 \vepsi_0 \partialt{\mbf{E}} \;, \\
\divr \mbf{B} = 0 \;,& &\curl \mbf{E} = - \partialt{\mbf{B}} \;,
\eea in Lorentz covariant notation as \bea \label{eqn-MaxF}
&\partial_\mu F^{\mu \nu} = (\partial_\mu \partial^\mu) A^\nu - \partial^\nu (\partial_\mu A^\mu) = \mu_0 J^\nu \;, \\
&\partial_\mu \wt{F}^{\mu \nu} = \partial^\lambda F^{\mu \nu} + \partial^\mu F^{\nu \lambda} + \partial^\nu F^{\lambda \mu} = 0 \;,
\eea where the field tensor is the four-curl of the electromagnetic potential, $F^{\mu \nu} \equiv \partial^\mu A^\nu - \partial^\nu A^\mu$, and its dual is $\wt{F}^{\mu \nu} \equiv \vepsi^{\mu \nu \alpha \beta} F_{\alpha \beta}/2$, where $\vepsi^{\mu \nu \alpha \beta}$ is the permutation tensor.  In natural units and intrinsic form, one writes $\dext\, ^*\!F = J$ and $\dext\, F = 0$ for $F = \dext\, A$, and Maxwell's theory amounts to the relation between the source current 3-form and the connection 1-form through the curvature 2-form, $\dext\, ^* \dext\, A = J$.  Current is conserved as a consequence of gauge invariance and Noether's theorem~[\cite{noether-235,tavel-183}], $\partial_\mu J^\mu = 0$, just as energy and momentum are conserved from space-time translational invariance.  Here we consider the neutral fluid limit $\rho_e \rightarrow 0$ such that $\divr \mbf{E} \rightarrow 0$, which distinguishes the EM-HD model from that of resistive magnetohydrodynamics (MHD) using the quasineutral approximation.  The EM-HD model is not restricted to the neutral fluid limit; however, the fluid equations of motion become much more complicated for a plasma with non-vanishing space charge density.

\subsection{Mass and energy equations}
The equations for mass and energy conservation are taken as the continuity equation with adiabatic closure, applicable by species, \beq \label{eqn-contin}
\ddt{n_s m_s} + \divr (n_s m_s \mbf{V}_s) = 0 \;,\; \DDt{}\frac{p_s}{(n_s m_s)^\gamma} = 0 \;,
\eeq where $D/Dt \equiv \partialt{} + \mbf{V}_s \cdot \del$ is the convective derivative and noting that the effect of plasma heating through resistive dissipation is not incorporated.  The degrees of freedom for the mass and energy density carried by a species are given by the species particle density and temperature.  Here, we account for these equations by assuming their degrees of freedom in the form of the density and temperature profiles \beq \label{eqn-profile}
n_s(r) = n_s(0) + \left[ n_s(r_1) - n_s(0) \right] (r/r_1)^{a_{ns}} \;,
\eeq and similarly for $\Tbar_s$, for radius $0 \leq r \leq r_1$ and profile exponent $a = 2$.

\subsection{Momentum equations}
The $\partialt{} \rightarrow 0$ equilibrium equations of motion, retaining from the stress tensor only the gradient of the scalar pressure, read \beq \label{eqn-eqnofmtn}
n_s m_s \left( \mbf{V}_s \cdot \del \right) \mbf{V}_s + \del p_s = n_s e_s \left( \mbf{E} + \mbf{V}_s \times \mbf{B} \right) + \mbf{F}_{s k} + \mbf{F}_{M s} \;,
\eeq where $\mbf{F}_{M s}$ is the magnetization force to be discussed later and $k\neq s$ for the friction term $\mbf{F}_{e i} = - \mbf{F}_{i e} = m_e \nu_{e i} \mbf{J}_f/e$ representing interspecies collisions.  From the definitions of the momentum and current densities, \beq
\left[\begin{array}{c} \rho_m \mbf{V}_f\\\mbf{J}_f \end{array} \right] = n_0 \left[\begin{array}{cc} m_i & m_e \\ e & -e \end{array} \right] \left[\begin{array}{c} \mbf{V}_i\\\mbf{V}_e \end{array} \right] \;,
\eeq we may exchange the ion and electron velocities $\{\mbf{V}_i, \mbf{V}_e\}$ for the pair $\{\mbf{V}_f, \mbf{J}_f\}$, \beq
\left[\begin{array}{c} \mbf{V}_i\\\mbf{V}_e \end{array} \right] = \mbf{V}_f + \frac{\mbf{J}_f}{e \rho_m} \left[\begin{array}{r} \ m_e\\-m_i \end{array} \right]\;.
\eeq  The sum of Equations~(\ref{eqn-eqnofmtn}) gives the equilibrium net force balance equation \beq \label{eqn-equileqn}
\mbf{C}_+ + \del p_+ - \mbf{F}_{M+} = \mbf{J}_f \times \mbf{B} \;,
\eeq and their difference the generalized Ohm's law equation for steady currents \beq \label{eqn-ohmseqn}
\mbf{C}_- + \del p_- - \mbf{F}_{M-} = 2(n_0 e \mbf{E} - \mbf{F}_{e i}) + [2 n_0 e \mbf{V}_f - n_0 (m_i - m_e) \mbf{J}_f / \rho_m] \times \mbf{B} \;,
\eeq where $p_\pm \equiv p_i \pm p_e$ and similarly for the magnetization force $\mbf{F}_{M\pm}$, and the convective terms $\mbf{C}_\pm \equiv n_0 \left[ m_i \left( \mbf{V}_i \cdot \del \right) \mbf{V}_i \pm m_e \left( \mbf{V}_e \cdot \del \right) \mbf{V}_e \right]$ are given by \bea \label{eqn-convterms}
\mbf{C}_+ =& n_0 \left(m_i + m_e\right) & \left[ \left( \mbf{V}_f \cdot \del \right) \mbf{V}_f + \frac{m_e m_i}{e^2} \left(\frac{\mbf{J}_f}{\rho_m} \cdot \del \right) \frac{\mbf{J}_f}{\rho_m} \right] ,\\
\mbf{C}_- =& n_0 \left(m_i - m_e\right) & \left[ \left( \mbf{V}_f \cdot \del \right) \mbf{V}_f - \frac{m_e m_i}{e^2} \left(\frac{\mbf{J}_f}{\rho_m} \cdot \del \right) \frac{\mbf{J}_f}{\rho_m} \right] \\ & + \; 2 n_0 m_e m_i & \left[ \left( \mbf{V}_f \cdot \del \right) \frac{\mbf{J}_f}{e \rho_m} + \left(\frac{\mbf{J}_f}{e \rho_m} \cdot \del \right) \mbf{V}_f \right] \;.
\eea  With a rearrangement of factors, the generalized Ohm's law equation may be put into the form \beq \label{eqn-myohm1}
\left[\mbf{C}_- + \del p_- - \mbf{F}_{M-}\right] / 2 n_0 e = \mbf{E} -  \eta \mbf{J}_f + \mbf{V}_f \times \mbf{B} - (m_i - m_e) \mbf{J}_f \times \mbf{B} / 2 e \rho_m \;,
\eeq using the resistivity $\eta=m_e \nu_{e i} / n_0 e^2$ for interspecies collision rate $\nu_{e i}$.  The final term in Equation~(\ref{eqn-myohm1}) has acquired a factor of 1/2 relative to its usual expression~[\cite{dendybook-93,sturrock}], which in our notation upon $m_e \ll m_i$ would equal $\mbf{J}_f \times \mbf{B} / n_0 e$.  The reason for the difference is that the standard derivation, with its application of extraneous mass factors to the equations of motion, does not respect the unwritten factor of units, as follows: given Equations~(\ref{eqn-eqnofmtn}) for the net species forces $\mbf{F}_i$ and $\mbf{F}_e$ in SI units of force-density, ${\rm Nt/m^3}$, the standard derivation would take the difference equation to be $m_e \mbf{F}_i - m_i \mbf{F}_e$ followed by $m_e \ll m_i$; however, the amount of force present did not change, and reinstating the unit factors reveals that $\mbf{F}_- ({\rm Nt/m^3}) = [m_e ({\rm kg}) \mbf{F}_i] ({\rm Nt/m^3/kg}/m_e) - [m_i ({\rm kg}) \mbf{F}_e] ({\rm Nt/m^3/kg}/m_i)$.  One must be careful to ensure that all one's quantities are in the same (SI) units before adding or subtracting them to achieve the physical result.  Following the standard derivation~[\cite{dendybook-93}], one would compare the numerator factor $(m_e+m_i)/m_i \rightarrow 1$ with $1+1=2$.  That factor is important, for it predicts that the effect of the Hall term on the driven current is only half of what is commonly taken, and upon neglect of the fluid velocity $\mbf{V}_f \rightarrow 0$, one may substitute the remainder of the net force balance Equation~(\ref{eqn-equileqn}) for the Hall term in the Ohm's law equation, whereupon neglect of the pressure gradients $\del p_{i,e} \rightarrow 0$ and any residual magnetization force $\mbf{F}_{M\pm} \rightarrow 0$ yields \beq \label{eqn-simplohm}
n_0 e \mbf{E} - \mbf{F}_{e i} = \half \left[ \mbf{C}_- + \frac{n_0 \left(m_i - m_e\right)}{\rho_m} \mbf{C}_+ \right] = 0\;,
\eeq giving $\mbf{J}_f = \sigma \mbf{E}$ for DC conductivity $\sigma = 1/\eta$.  Reinstating the acceleration term $\propto \partlxy{\mbf{J}_f}{t}$ recovers the AC conductivity~[\cite{rwj-pop01}] and with vanishing damping factor a plasma frequency of $\omega_p = e \sqrt{n_0 (m_i + m_e)/m_i m_e \vepsi_0}$.
  Equilibrium requires that neither the momentum nor the current diverge, $\divr (\rho_m \mbf{V}_f) = \divr \mbf{J}_f = 0$.  We next account for gyromotion by the substitutions $\mbf{V}_f \rightarrow \mbf{V}_f + \mbf{V}_d = \mbf{V}$ and $\mbf{J}_f \rightarrow \mbf{J}_f + \mbf{J}_d = \mbf{J}$ everywhere except the friction term $\mbf{F}_{e i}$ representing collisional disruption of the free motion, as the collisional disruption of the gyromotion will be considered within the nonlinear magnetization model.

\subsection{Plasma magnetization}
Consider coaxial applied electric and magnetic fields $\mbf{E}_0$ and $\mbf{H}_0$.  The electric field will drive a free current $\mbf{J}_f$ that in turn creates a magnetic field $\mbf{H}_f$ which in conjunction with the applied field $\mbf{H}_0 + \mbf{H}_f = \mbf{H}$ produces a magnetization $\mbf{M}$ resulting from the gyromotion of the constituent particles. Our approach to the treatment of magnetization most closely follows that found in References~[\cite{hazeltine-04,griffiths-89,kauf-9212}], except that we will be fully decomposing the Hall term $\mbf{J} \times \mbf{B}$ in terms of $\mbf{H}$ and $\mbf{M}$, and our logic follows \beq
\mbf{E}_0 \rightarrow \mbf{J}_f \rightarrow \mbf{H}_f + \mbf{H}_0 \rightarrow \mbf{M} \;,
\eeq from which one finds the drift velocity and the diamagnetic current from the curls of the net gyro-momentum and magnetic dipole moment densities, respectively: $\rho_m \mbf{V}_d \equiv \curl \mbf{L}_g$ and $\mbf{J}_d \equiv \curl \mbf{M}$.  The gyrovector is defined by $\vec{\omega}_s \equiv - e_s \mbf{B}_s / m_s$, where the field felt by a particle of species $s$ is $\mbf{B}_s/\mu_0 = \mbf{H} + \mbf{M} - \vec{\mu}_s$ and points along $\bhat_s \equiv \hhat \equiv \mbf{H}/H$ for $\vec{\mu}_s$ the magnetic moment of a single particle.  The net gyro-momentum is the sum of each species contribution, $\mbf{L}_g = \sum_s f_s n_s \vec{l}_s$, and similarly for the net magnetization, $\mbf{M} = \sum_s f_s n_s \vec{\mu}_s$, where the particle gyro-momentum is the cross product of the gyroradius $r_{g s} = v^\perp_s / \omega_s$ and the perpendicular momentum, $\vec{l}_s = \mbf{r}_{gs} \times m_s \mbf{v}^\perp_s = 2 \Tbar_s \vec{\omega}_s / \omega_s^2$, the magnetic dipole moment is $\vec{\mu}_s = - (W^\perp_s / B_s) \bhat_s = - \Tbar_s \mbf{B}_s / B_s^2$, and the factor $f_s$ representing collisional disruption of the gyromotion is modelled as \beq \label{eqn-fs}
f_s = \frac{\omega_s}{\nu_s} \left(1 - e^{- \nu_s / \omega_s} \right) \left \lbrace \begin{array}{lc} \rightarrow 1 & \mathrm{for}\; \omega_s \gg \nu_s \;, \\ \rightarrow \omega_s / \nu_s & \mathrm{for}\; \omega_s \ll \nu_s \;, \end{array} \right. \;,
\eeq where $\nu_s \equiv \nu_{s s} + \nu_{s k}$ is the net species collision rate, and is normalized so that $f p = \sum_s f_s p_s$.  Scattering times $\tau_{s s'} \equiv 1/ \nu_{s s'}$ are calculated using common formulas~[\cite{tokamaks-2004,physics-nrl}] as $\tau_{e e} = C \vepsi_0^2 m_e^{1/2} \Tbar_e^{3/2} / n_e e^4 \ln \Lambda_{e e}$, $\tau_{e i} = C \vepsi_0^2 m_e^{1/2} \Tbar_e^{3/2} / \sqrt{2} n_i e_i^2 e^2 \ln \Lambda_{e i}$, $\tau_{i i} = C \vepsi_0^2 m_i^{1/2} \Tbar_i^{3/2} / n_i e_i^4 \ln \Lambda_{i i}$, and $\nu_{i e} = n_e m_e \nu_{e i} / n_i m_i$, for constant $C = 12 \pi^{3/2}$.  With plasma parameter $\alpha_s \equiv (n_s - 1) / n_s$ approximately unity for sufficient density, the species field is $\mbf{B}_s / \mu_0 = \mbf{H} + \mbf{M}_k + \alpha_s \mbf{M}_s$, and using $\ptild \equiv p / \mu_0$ gives a final magnetization model for $\mbf{M} = - M \hhat = 2 n_0 (f_e \vec{\mu}_e + f_i \vec{\mu}_i)/2$ of \beq
M = \frac{f_e \ptild_e}{H-M_i - \alpha_0 M_e} + \frac{f_i \ptild_i}{H-M_e- \alpha_0 M_i} = \frac{f \ptild}{H - \alpha M} \;,
\eeq which has solution $M/H = (1 - \sqrt{1 - 4 \alpha f \ptild / H^2})/2 \alpha$ for $0 < \alpha \leq 1$ and $\beta$-limit $\beta \equiv 2 \ptild / H^2 \leq 1/ 2 \alpha f$ on the ratio of the kinetic to the magnetic pressure, which goes to $1/2$ for $\alpha, f \rightarrow 1$ and to $1/f$ for $\alpha \rightarrow 1/2$ as $\alpha_0 \rightarrow 0$.  As $\omega_s$ is in terms of $\mbf{B}_s$, an iterative approach to the collisionality factor may be defined by $M_n (H-\alpha M_n) = f_n \ptild$, starting at $f_0 = 1$ with $f_n = f_n(M_{n-1})$, and for a dense, magnetically confined plasma we find $f$ remains very close to unity.  The decomposed Hall term then reads, as $\mbf{J}_0 = \curl \mbf{H}_0$ is nonzero only for $r > r_1$, \bea \label{eqn-Hallterm}
\mbf{J} \times \mbf{B} / \mu_0 &=& \left[ \curl (\mbf{H} + \mbf{M}) \right] \times (\mbf{H} + \mbf{M}) \;, \\
 &=& \left[ (\mbf{H}+\mbf{M}) \cdot \del \right] (\mbf{H}+\mbf{M}) - \del | \mbf{H}+\mbf{M} |^2 / 2 \;, \\
 &=& \left[ (\mbf{H}+\mbf{M}) \cdot \del \right] (\mbf{H}+\mbf{M}) - H \del H - M \del M + \del M H \;.
\eea

\subsection{Magnetization force}
The magnetization of the plasma medium also gives rise to a magnetization force felt by the dipoles in an external field~[\cite{griffiths-89,feynmanlecs}].  Here we consider a generalization of the macroscopic force densities~[\cite{melcher-81,rosen-82}] given by Lorentz and Kelvin, $\mbf{F}_\mathrm{LK} = \mu_0 \mbf{J} \times \mbf{H} + \mu_0 \mbf{M} \cdot \del \mbf{H}$, and by Korteweg and Helmholtz, $\mbf{F}_\mathrm{KH} = \mbf{J} \times \mbf{B} - \mbf{H} \cdot \mbf{H} \del \mu / 2$, modelled as \beq
\mbf{F}_M \equiv \mu_0 \del \mbf{M} \cdot \mbf{H} = - \mu_0 \del M H \;,
\eeq for $\mbf{M} = - (M/H) \mbf{H}$, noting that its presence ensures that the Hall term of Equation~(\ref{eqn-Hallterm}) reduces to the correct form $\mbf{J}_f \times \mbf{H} = (\mbf{H} \cdot \del) \mbf{H} - H \del H$ in the free-current limit $J_d \ll J_f$~[\cite{rwj-ppcf03}].  We consider this force an important effect neglected in the usual analysis of plasma equilibrium despite its experimental applications in fusion~[\cite{hayes-213}], magnetic fluids~[\cite{rinaldi-2847,zahn-144,eldib-159}], biophysics~[\cite{tzir-077103,rama-297,qi-132,ataka-ic0038,wang-877}], and materials science~[\cite{maki-1132,maki-1096,maki-066106,ma-2944,asai-r1,takagi-842,kozuka-884,fangwei-024202,colli-58,ono-608,cantor-02,keil-07}] and addressed for the case of a stationary equilibrium elsewhere~[\cite{rwj-ppcf03}].  In the evaluation of Equations~(\ref{eqn-equileqn}) and (\ref{eqn-ohmseqn}), we take $\mbf{F}_{M\pm} \equiv \mbf{F}_{M i} \pm \mbf{F}_{M e}$ where $\mbf{F}_{M s} = - \mu_0 \del M_s H$.  There remains to incorporate any effects arising from the intrinsic spin of the particles, requiring a properly quantum mechanical treatment of plasma magnetization.

\section{Numerical evaluation}
\subsection{Restricted equations}
With restriction to an infinite $\partial / \partial z \rightarrow 0$ plasma column of meter radius $r_1 = 1$ with azimuthal symmetry $\partial / \partial \theta \rightarrow 0$, equilibrium requires that $V_{f r} = J_{f r} = 0$.  The applied coaxial fields are taken as $\mbf{E}_0 = E_0 \zhat$ and $\mbf{H}_0 = H_0 \zhat$.   Then, the fluid equations of motion, Equations~(\ref{eqn-equileqn}) and (\ref{eqn-ohmseqn}), for a sufficiently dense $\alpha \rightarrow 1$ medium, reduce to a system of four scalar equations in four unknowns, $\{V_{f \theta}, V_{f z}, J_{f \theta}, J_{f z}\}$.  The free current is found from the $\theta$ and $z$ components of the Ohm's law equation, which in this case gives us $J_{f \theta} = 0$ and $J_{f z} = n_0 e^2 E_0 / m_e \nu_{ei}$.  The enclosed free current as a function of radius $I_f(r) = \int_0^r 2 \pi r' J_{f z}(r') dr'$ then determines the azimuthal magnetic field $\mbf{H}_f = H_f \thetahat = (I_f / 2 \pi r) \thetahat$.  Five iterations were found sufficient to converge the collisionality factor $f$ in the determination of the magnetization $\mbf{M}$, and the drift current found from its curl is \beq \label{eqn-curlM}
\mbf{J}_d \equiv \curl \mbf{M} = - \curl \left( \MoH \mbf{H} \right) = \mbf{H} \times \del \MoH - \MoH \curl \mbf{H} \;,
\eeq and similarly for the drift momentum $\rho_m \mbf{V}_d \equiv \curl \mbf{L}_g$, from which in $(r,\theta,z)$ coordinates \beq \label{eqn-Jd}
\mbf{J}_d = \left(0, H_0 \ddr{} \MoH, -H_f \ddr{} \MoH - J_z \MoH \right) \;.
\eeq  The net magnetic field is given by $\mbf{B} = \mu_0 (1-M/H)\mbf{H}$.  We next consider the solution of the remaining radial equations for both the free current model $\mbf{V}^f_f$, valid when $\ptild \ll H^2$ such that $M \ll H$ and $J_d \ll J_f$, and the magnetized model $\mbf{V}^m_f$ with $\mbf{V}^m = \mbf{V}^m_f + \mbf{V}^m_d$.  The free current model has the solution \bea \label{eqn-freesoln}
V^f_{f \theta} &=& \sqrt{\left(\ddr{p_+} + \mu_0 J_{f z} H_f \right) \frac{r}{\rho_m}} \;, \\
V^f_{f z} &=& \oover{2 n_0 e \mu_0 H_f} \left[\frac{n_0 (m_i - m_e)}{r} {V^f_{f \theta}}^2 - \ddr{p_-} \right] + \frac{(m_i - m_e)}{2 e \rho_m} J_{f z} + \frac{H_0}{H_f} V^f_{f \theta} \;,
\eea
%
and the magnetized model has the more complicated solution \bea \label{eqn-magsoln}
V^m_{f \theta} &=& \sqrt{\left(\ddr{p_+} - F_{M+} + J_z B_\theta - J_\theta B_z \right) \frac{r}{\rho_m} - \frac{m_e m_i}{e^2 \rho_m^2} J_\theta^2} - V^m_{d \theta} \;, \\
V^m_{f z} &=& \oover{2 n_0 e B_\theta} \left[\frac{n_0 (m_i - m_e)}{r} \left( {V^m_\theta}^2 - \frac{m_e m_i}{e^2 \rho_m^2} J_\theta^2 + \frac{4 m_e m_i / e \rho_m}{m_i - m_e} V^m_\theta J_\theta \right) - \ddr{p_-} + F_{M-} \right] \\
 & & + \frac{(m_i - m_e)}{2 e \rho_m} \left(J_z - \frac{B_z}{B_\theta} J_\theta \right) + \frac{B_z}{B_\theta} V^m_\theta  - V^m_{d z} \;.
\eea

\subsection{Comparative solutions}
\begin{figure}
\includegraphics*[bb = 91 135 533 544, scale=.9]{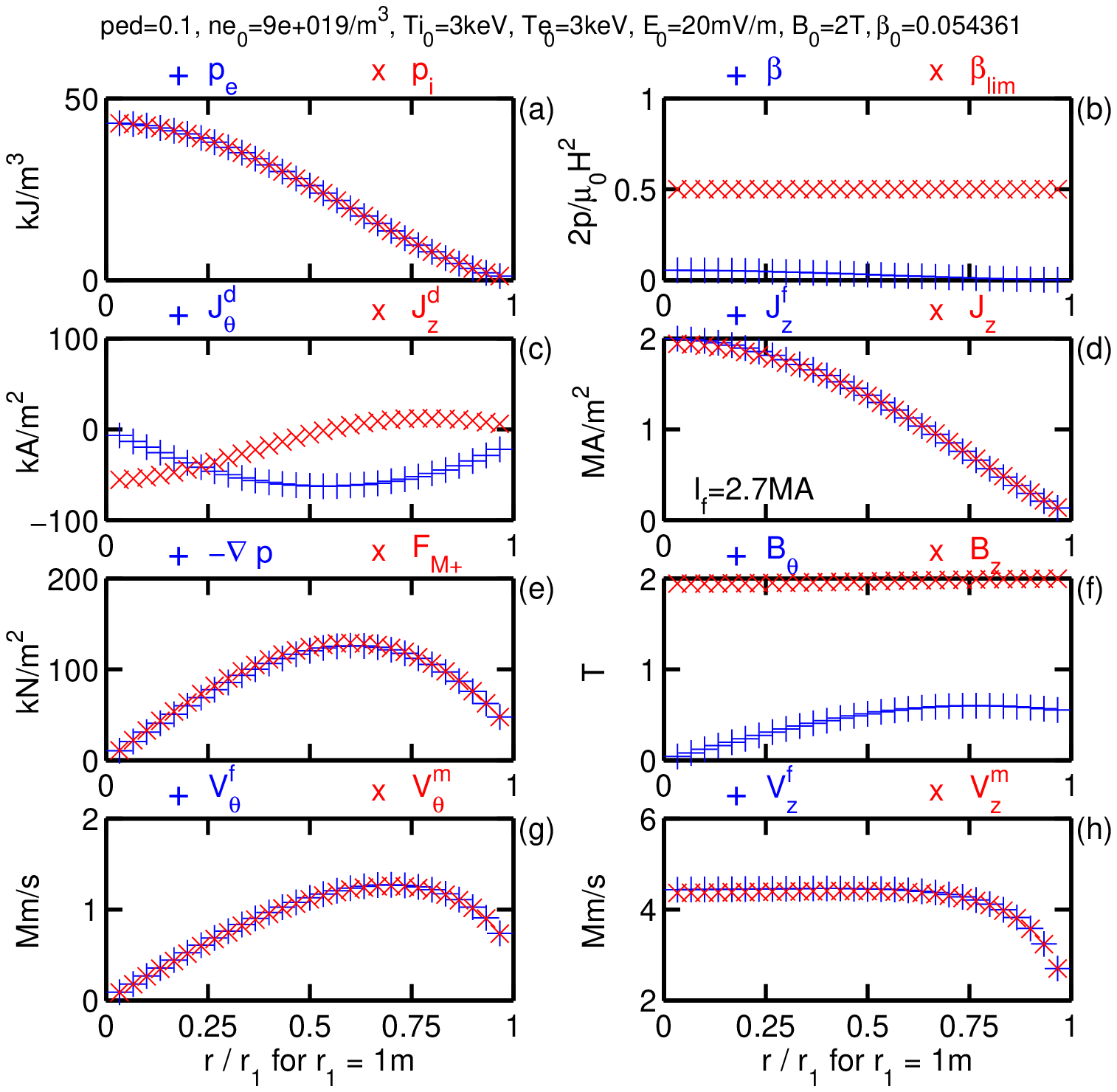}
\caption{\label{figA}(Color online.)  Solution profiles for $\chi_p = 1/10$, $n_0 = 9 \times 10^{19}/{\rm m}^3$, $\Tbar_e = 3$keV, $\Tbar_i = 3$keV, $E_0 = 20$mV/m, $B_0 = 2$T, and $\beta_0 = 5.4\%$. (a) Electron pressure $+$ and ion pressure $\times$. (b) Kinetic to magnetic pressure ratio $\beta\;+$ and its limit $\beta_{\rm lim}\;\times$. (c) Diamagnetic current in azimuthal $+$ and axial $\times$ directions. (d) Axial free current $+$ and net current $\times$. (e) Gradient forces for pressure $- \del p \; +$ and magnetization $F_{M} \; \times$. (f) Net magnetic field in the azimuthal $+$ and axial $\times$ directions. (g) Azimuthal fluid velocity for the free current model $+$ and the magnetized model $\times$. (h) Axial fluid velocity for the free current model $+$ and the magnetized model $\times$.}
\end{figure}

In the following we consider a plasma with central density $n_0(0) = 9 \times 10^{19}/{\rm m}^3$ and central electron temperature $\Tbar_e(0) = 3$keV immersed in an electric field $E_0 = 20$mV/m and magnetic field $B_0 = 2$T.  We will compare the solutions for a central ion temperature $\Tbar_i$ of 3keV and 12keV (``cold'' and ``hot''), where the outer ion temperature is set equal to that of the electrons, $\Tbar_i(r_1) = \Tbar_e(r_1)$, first for a pedestal ratio $\chi_n \equiv n_0(r_1)/n_0(0) = 1/10$ and then for a ratio of $1/\sqrt{10}$ (and similarly for $\Tbar_e$).  The pressure pedestal ratio $\chi_p$ goes as the product of the density and temperature pedestal ratios.  For the first solution with cold ions and $\chi_p = 1/100$, Figure~\ref{figA}, we find that the central $\beta_0 \approx 5.4\%$ remains well below the limit of 50\% and that the diamagnetic current is on the order of $\rm kA/m^2$.  The drift current's axial component serves both to suppress and enhance the net axial current relative to the free axial current, and the total free current for this configuration is approximately 2.7MA.  The net magnetization force is found to equal or exceed the pressure gradient force across the profile, and the net axial magnetic field reflects the diamagnetic contribution.  The net fluid velocity is on the order of Mm/s, with the maximum axial component exceeding the maximum azimuthal component by a ration of 4/1, and the drift velocity is on the order of mm/s.  The axial fluid velocity for the magnetized model is slightly suppressed near the core compared to the free current model.

\begin{figure}
\includegraphics*[bb = 91 135 533 544, scale=.9]{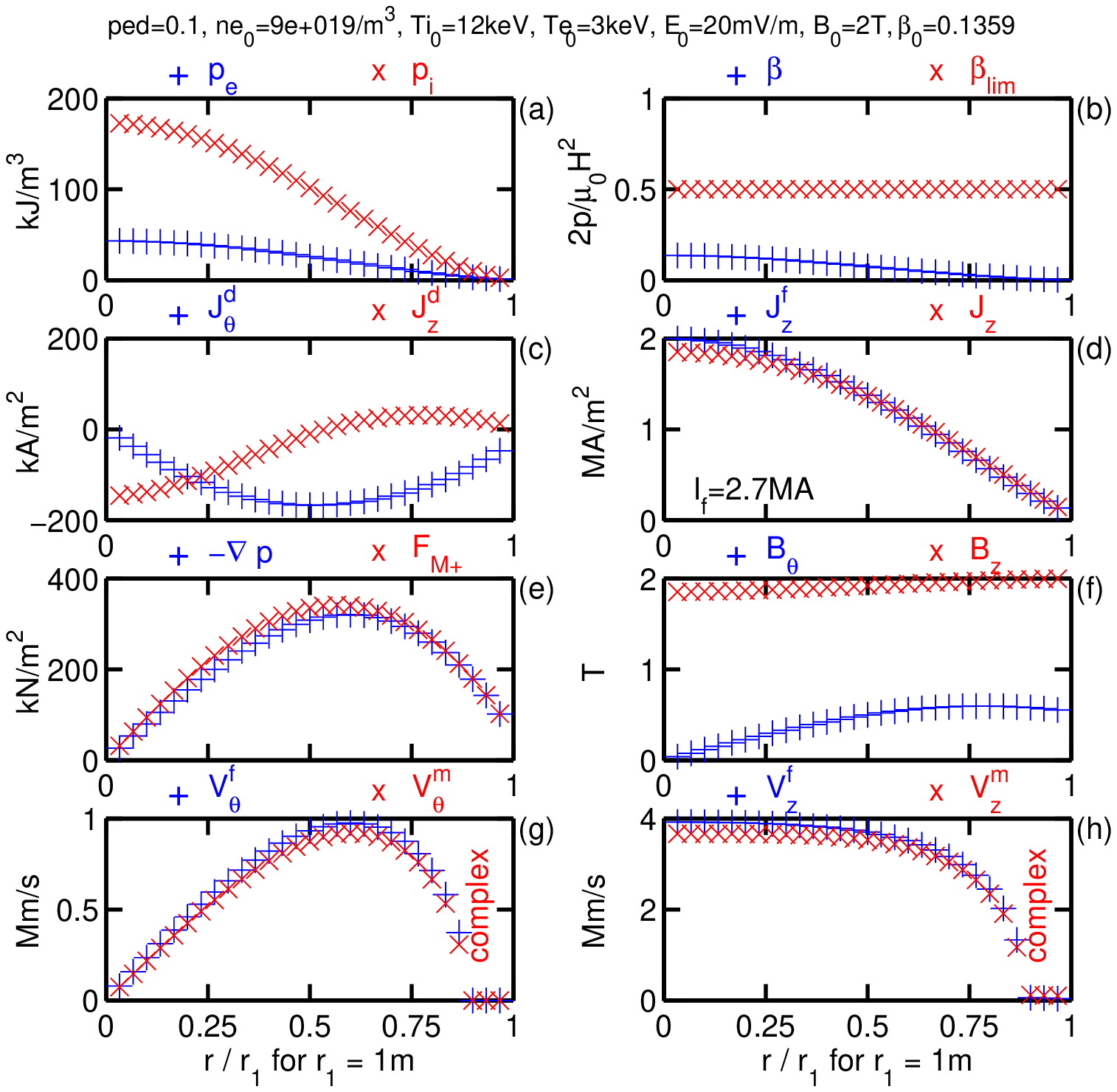}
\caption{\label{figB}(Color online.)  Solution profiles for $\chi_p = 1/10$, $n_0 = 9 \times 10^{19}/{\rm m}^3$, $\Tbar_e = 3$keV, $\Tbar_i = 12$keV, $E_0 = 20$mV/m, $B_0 = 2$T, and $\beta_0 = 13.6\%$. (a) Electron pressure $+$ and ion pressure $\times$. (b) Kinetic to magnetic pressure ratio $\beta\;+$ and its limit $\beta_{\rm lim}\;\times$. (c) Diamagnetic current in azimuthal $+$ and axial $\times$ directions. (d) Axial free current $+$ and net current $\times$. (e) Gradient forces for pressure $- \del p \; +$ and magnetization $F_{M} \; \times$. (f) Net magnetic field in the azimuthal $+$ and axial $\times$ directions. (g) Azimuthal fluid velocity for the free current model $+$ and the magnetized model $\times$. (h) Axial fluid velocity for the free current model $+$ and the magnetized model $\times$.  Note that the solutions for the fluid velocity have been driven complex near the outer edge of the plasma column.}
\end{figure}

\begin{figure}
\includegraphics*[bb = 91 135 533 544, scale=.9]{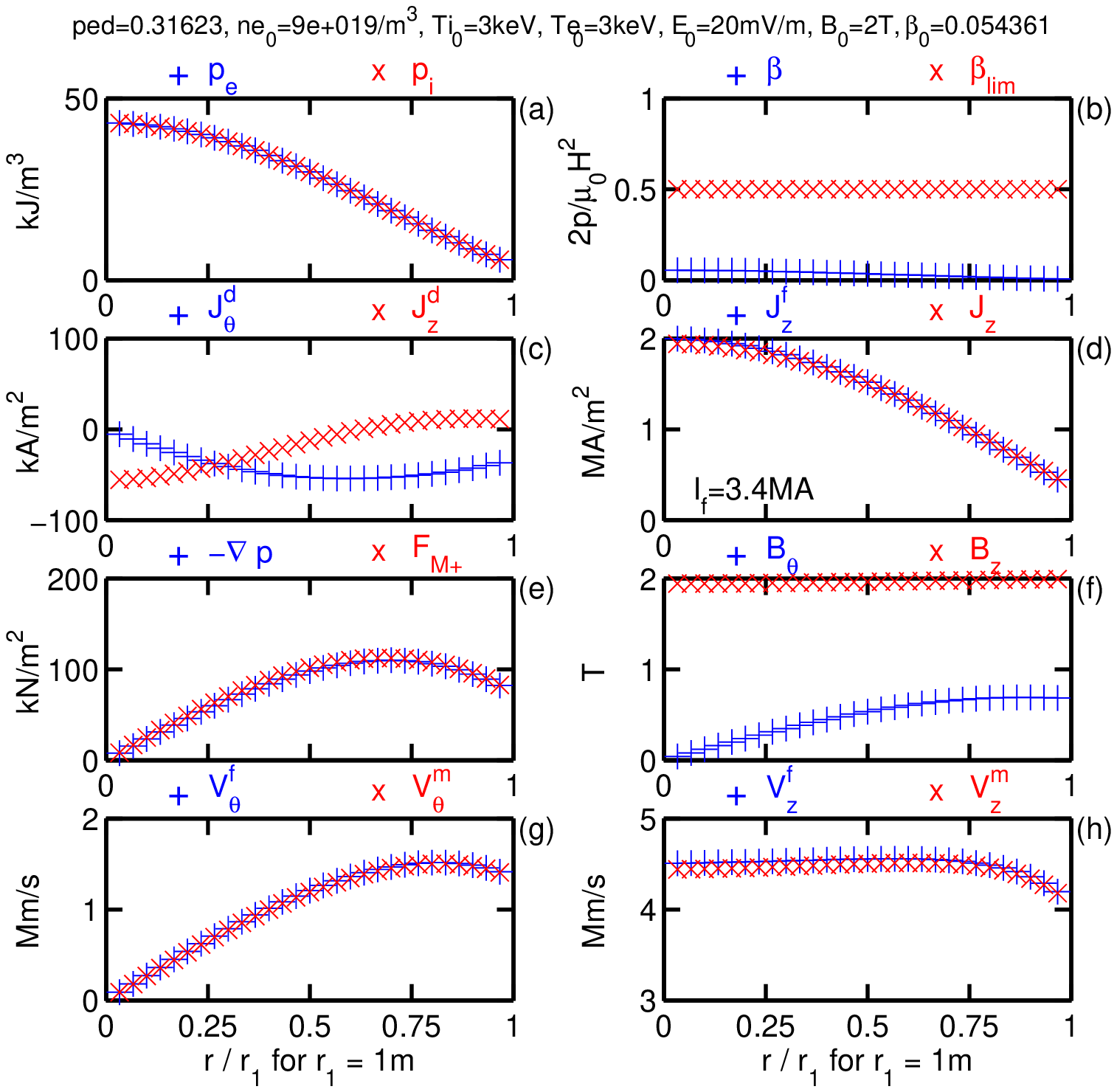}
\caption{\label{figC}(Color online.)  Solution profiles for $\chi_p = 1/\sqrt{10}$, $n_0 = 9 \times 10^{19}/{\rm m}^3$, $\Tbar_e = 3$keV, $\Tbar_i = 3$keV, $E_0 = 20$mV/m, $B_0 = 2$T, and $\beta_0 = 5.4\%$. (a) Electron pressure $+$ and ion pressure $\times$. (b) Kinetic to magnetic pressure ratio $\beta\;+$ and its limit $\beta_{\rm lim}\;\times$. (c) Diamagnetic current in azimuthal $+$ and axial $\times$ directions. (d) Axial free current $+$ and net current $\times$. (e) Gradient forces for pressure $- \del p \; +$ and magnetization $F_{M} \; \times$. (f) Net magnetic field in the azimuthal $+$ and axial $\times$ directions. (g) Azimuthal fluid velocity for the free current model $+$ and the magnetized model $\times$. (h) Axial fluid velocity for the free current model $+$ and the magnetized model $\times$.}
\end{figure}

\begin{figure}
\includegraphics*[bb = 91 135 533 544, scale=.9]{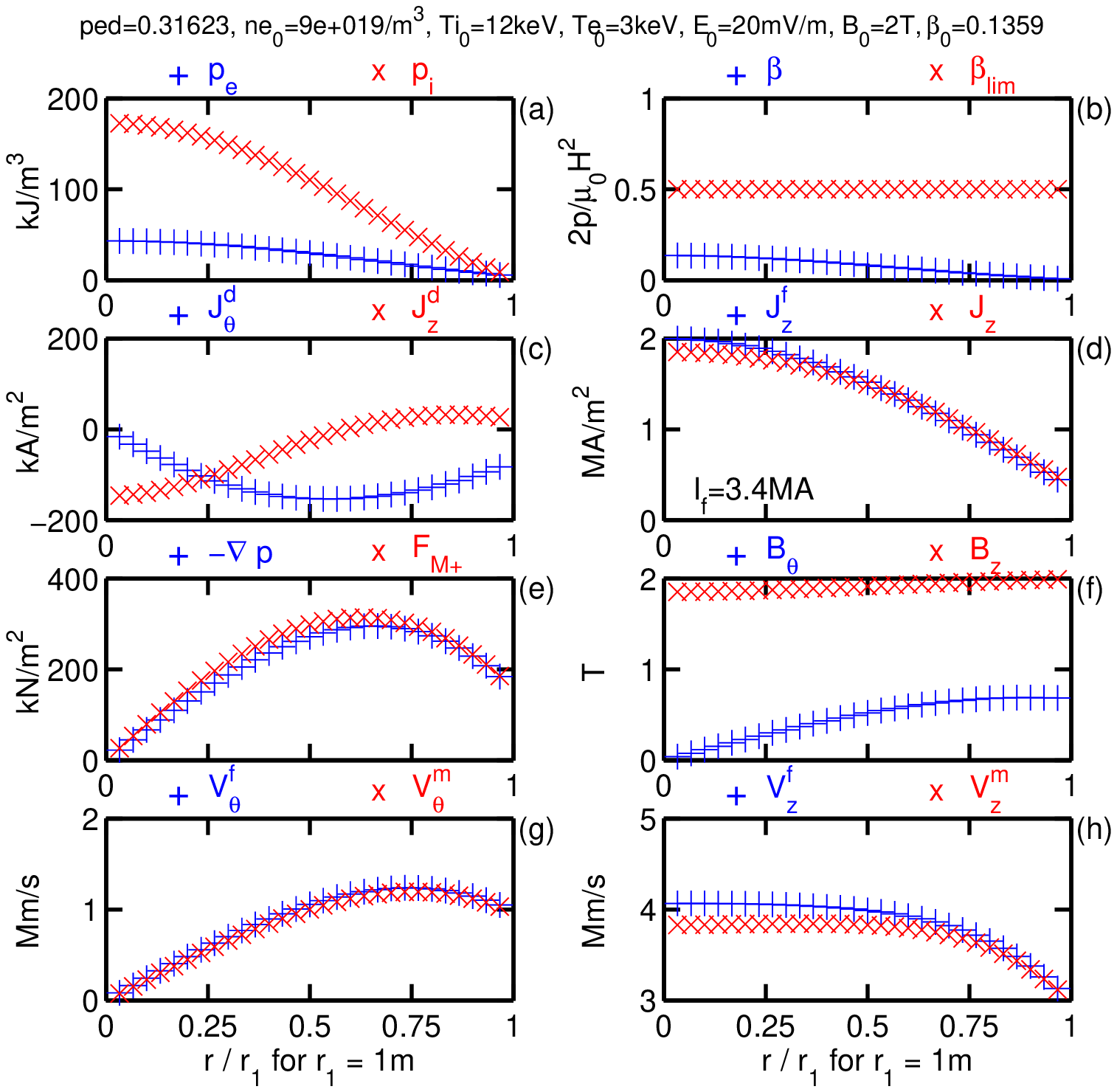}
\caption{\label{figD}(Color online.)  Solution profiles for $\chi_p = 1/\sqrt{10}$, $n_0 = 9 \times 10^{19}/{\rm m}^3$, $\Tbar_e = 3$keV, $\Tbar_i = 12$keV, $E_0 = 20$mV/m, $B_0 = 2$T, and $\beta_0 = 13.6\%$. (a) Electron pressure $+$ and ion pressure $\times$. (b) Kinetic to magnetic pressure ratio $\beta\;+$ and its limit $\beta_{\rm lim}\;\times$. (c) Diamagnetic current in azimuthal $+$ and axial $\times$ directions. (d) Axial free current $+$ and net current $\times$. (e) Gradient forces for pressure $- \del p \; +$ and magnetization $F_{M} \; \times$. (f) Net magnetic field in the azimuthal $+$ and axial $\times$ directions. (g) Azimuthal fluid velocity for the free current model $+$ and the magnetized model $\times$. (h) Axial fluid velocity for the free current model $+$ and the magnetized model $\times$.  Note that the slight increase in pressure at the outer edge alleviates the difficulty with the fluid velocity profile.}
\end{figure}

\begin{figure}
\includegraphics*[bb = 91 135 533 544, scale=.9]{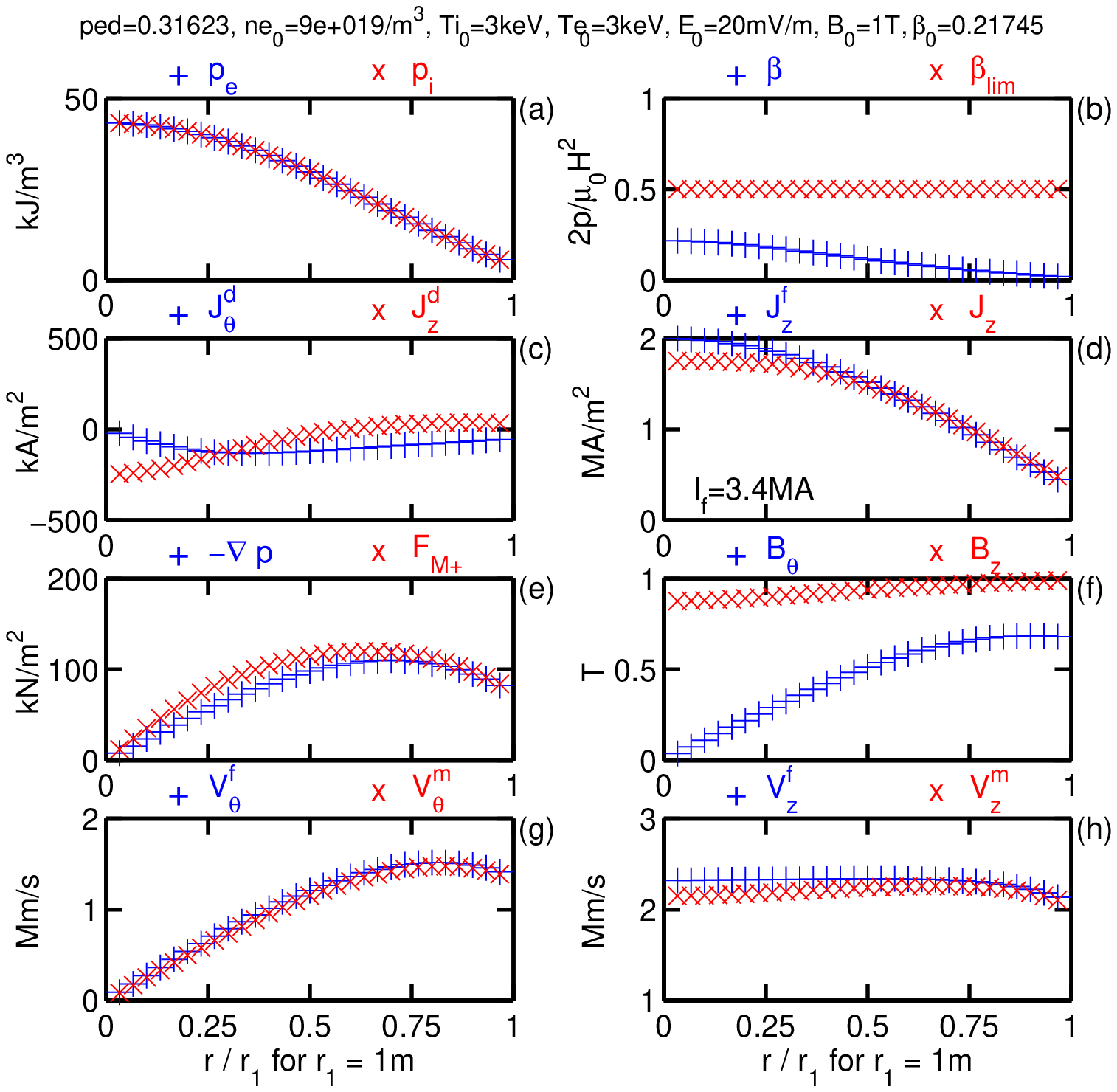}
\caption{\label{figE}(Color online.)  Solution profiles for $\chi_p = 1/\sqrt{10}$, $n_0 = 9 \times 10^{19}/{\rm m}^3$, $\Tbar_e = 3$keV, $\Tbar_i = 3$keV, $E_0 = 20$mV/m, $B_0 = 1$T, and $\beta_0 = 21.7\%$. (a) Electron pressure $+$ and ion pressure $\times$. (b) Kinetic to magnetic pressure ratio $\beta\;+$ and its limit $\beta_{\rm lim}\;\times$. (c) Diamagnetic current in azimuthal $+$ and axial $\times$ directions. (d) Axial free current $+$ and net current $\times$. (e) Gradient forces for pressure $- \del p \; +$ and magnetization $F_{M} \; \times$. (f) Net magnetic field in the azimuthal $+$ and axial $\times$ directions. (g) Azimuthal fluid velocity for the free current model $+$ and the magnetized model $\times$. (h) Axial fluid velocity for the free current model $+$ and the magnetized model $\times$.  Note that the axial fluid velocity has decreased by a similar factor of 2 compared to the $B_0 = 2$T profile.}
\end{figure}

For the hot ions with the same $\chi_p$, Figure~\ref{figB}, while the kinetic pressure is well within the limit, $\beta_0 \approx 13.6\%$, the solution for the azimuthal fluid velocity, both free current and magnetized model, has been driven complex near the outer edge of the column, taking the axial velocity along with it---this equilibrium is unfeasible.  Increasing the outer pressure so that $\chi_p = 1/10$ alleviates the difficulty for both the cold and hot ion configurations, Figures~\ref{figC} and \ref{figD} respectively, with a total free current about 3.4MA.  Reducing the applied magnetic field to $B_0 = 1$T for this $\chi_p$ returns a feasible solution for the cold ion configuration with $\beta_0 \approx 21.7\%$, Figure~\ref{figE}, which displays a marked reduction in the axial fluid velocity (but not the the azimuthal fluid velocity) relative to the higher field case.  For this field strength, the hot ions are simply too hot, with a central $\beta_0 \approx 54.4\%$ in excess of the limit.  The magnetization has been driven complex in the plasma core, affecting every quantity dependent upon it and leading to an unphysical solution.  Note that the quantities pertinent to the free current model are not affected by this calamity, indicating that the models may be distinguished for a suitably designed experiment.

\section{Conclusions and outlook}\label{sec:conc}

The equilibrium equations of motion for an axially symmetric, magnetized, hydrogenic plasma column in applied electric and magnetic fields in the neutral fluid limit are investigated.  Respecting the units for the net difference of the species forces results in a factor of 1/2 on the Hall term in the Ohm's law equation relative to its usual value.  The equations of motion, including the macroscopic magnetization force, are used to determine the free momentum and current densities from prescribed species density and temperature profiles, and Maxwell's theory is used to determine the electromagnetic fields from the source charge-current density in the neutral fluid limit.  The diamagnetic current and drift velocity are found from the net magnetization and gyro-momentum using a nonlinear magnetization model including a collisionality correction factor $f$.  The ratio of kinetic to (free) magnetic pressure is found to be limited by a $\beta_{\rm lim}$ which ranges from $1/f \gtrsim 1$ for a sparse plasma to $1/2 f \approx 1/2$ for a dense, magnetically confined plasma.

The restricted equations yield analytic solutions for the fluid flow and current which are computed for a variety of parameter profiles for both the free current and fully magnetized models.  The pressure pedestal ratio is found to affect the feasibility of the equilibrium through the fluid velocity near the outer edge of the column, and comparison of the solutions for cold $\Tbar_i = \Tbar_e$ and hot $\Tbar_i > \Tbar_e$ ion temperature profiles indicates that the attainable central pressure is given by the limit on the plasma magnetization.  The free current model neglecting magnetization is distinguished by having no pressure limit.

The EM-HD model given above is ripe for extension in several different directions.  A primary difficulty to overcome is the reduction of the stress tensor to the gradient of the scalar pressure, rather than a gyrotropic or gyroviscous tensor more appropriate for a magnetized medium of free charges~[\cite{brag-1965,frc-pop-2006}].  Reinstating the acceleration terms in the equations of motion would yield the EM-HD dielectric tensor, and incorporation of a dielectric or conductive boundary material would yield a model more descriptive of actual devices of fusion~[\cite{1995PhPl....2.2236C}] and propulsion~[\cite{aiaa08-4926,ispc08-0236}] interest.  Evaluating the theory for a $\del = (\partlxy{}{r},\partlxy{}{\theta},\partlxy{}{\phi}\rightarrow 0)$ geometry corresponds to a model for a tokamak, and reallowing $\partlxy{}{\phi}$ could be used for stellerator analysis.  Lifting the restrictions of the neutral fluid limit requires readdressing the force balance equations for non-vanishing space charge density, with an aim towards a manifestly covariant description of Maxwell-Minkowsky electrodynamics appropriate for a fluid of free charges.


\section{Appendix} \label{app:}

The resistive magnetohydrodynamic (MHD) equations as usually defined in the quasineutral approximation refer to a system of 14 scalar equations in 14 scalar variables, hence are determined to be complete and soluble.  These equations are a combination of Navier-Stokes and a subset of Maxwell's.  However, one of the vector equations is actually an identity when viewed from the potential formulation of electrodynamics, hence does not determine any degrees of freedom.  Only by reinstating Gauss's law does the system of equations become closed, allowing for the prediction of both the current and momentum from the equations of motion.

Many authors~[\cite{dendybook-93,staceybook05,chen-84,dinkbook-05,kivel95,spsbds03,mbk04,goldruth95}] define the low frequency resistive MHD equations as the zeroth and first order moments of the Vlasov equation with adiabatic closure in conjunction with the two curl equations among Maxwell's.  For the neutral fluid, the sum and difference of the ion and electron equations of motion give the net force balance equation and the generalized Ohm's law.  Using $D/Dt \equiv \partial / \partial t + \mbf{V}_f \cdot \del$, we write the usual equations (which do not distinguish between $\mbf{J}_f$ and $\mbf{J}_d$ nor $\mbf{H}$ and $\mbf{M}$): \bea \label{eqn:1}
\ddt{\rho_m} + \divr (\rho_m \mbf{V}_f) = 0 \;, & & \DDt{p/ \rho_m^\gamma} = 0 \;, \\ \label{eqn:2}
\rho_m \DDt{\mbf{V}_f} = \mbf{J} \times \mbf{B} - \del p \;, & & \eta \mbf{J} = \mbf{E} + \mbf{V}_f \times \mbf{B}\;, \\ \label{eqn:3}
\curl \mbf{E} = - \ddt{\mbf{B}} \;, & & \curl \mbf{B} = \mu_0 \mbf{J}\;,
\eea where $\eta$ is the resistivity and $\gamma$ is the appropriate index for the case under consideration, and the degrees of freedom are pressure $p$, mass density $\rho_m$, flow velocity $\mbf{V}_f$, current $\mbf{J}$, and electromagnetic fields $\mbf{E}$ and $\mbf{B}$, giving a naive counting of 14 scalar equations for 14 scalar variables.  However, while for decades~[\cite{brag-1965,roseclark}] the argument has been made that Gauss's law may be neglected with impunity, no one within the plasma physics community has denied the applicability of the potential formulation of electrodynamics~[\cite{maxwell-1864}].  The unnamed of Maxwell's equations (often called the ``no-monopole'' equation, $\divr \mbf{B} = 0$) is brought into play during the determination of plasma equilibrium, via solution of the Grad-Shafranov equation~[\cite{grad-58,shafranov-66}] in toroidal geometry or otherwise, which by the naive counting of above would introduce an additional scalar equation, thus over-determining the system, yet is commonly known simply to allow for the expression of the magnetic field in terms of the vector potential, $\mbf{B} = \curl \mbf{A}$.  The reason doing so is valid is because vector identities by mathematical definition do not determine any degrees of freedom; they reduce them.  Inserting that expression into Faraday's law~[\cite{griffiths-89}], we recover $\curl (\mbf{E} + \partial \mbf{A} / \partial t) = 0$, whence $\mbf{E} = -\del \Phi - \partial \mbf{A} / \partial t$, which clearly displays the division of the electric field into static and dynamic components and reduces three of our naive degrees of freedom down to one for which we have no equation.  Unless one wishes to invent new physics, the resolution is clear---the reinstatement of Gauss's law, $\divr \mbf{E} = - \del^2 \Phi - \partial (\divr \mbf{A})/\partial t = \rho_e/\vepsi_0$ which vanishes for a neutral fluid, is required to close the system of equations, bringing the number of scalar equations and degrees of freedom into agreement with the number $14-3+1=14-2=12$.  We remark that Faraday's law is no less an identity than the no-monopole equation as both are given by the general theory of vector fields.  Gauge invariance plays a special role in the local conservation of charge, best expressed in manifestly Lorentz covariant notation.

From a particle physicist's field-theoretic point of view~[\cite{halzenmartin,ryder-qft,davis70,ramond-FTM90,mandlshaw}], the Maxwell field tensor $F^{\mu \nu} \equiv \partial^\mu A^\nu - \partial^\nu A^\mu$ in media is known to have only 3 physical degrees of freedom embodied by the four-potential $A^\mu \equiv (\Phi/c,\mbf{A})$ subject to the gauge condition, not 3 for each of the electric and magnetic fields, which couple to sources given by the conserved four-current $J^\mu \equiv (c \rho_e,\mbf{J})$ through the inhomogeneous Maxwell equations $\partial_\mu F^{\mu \nu} = (\partial_\mu \partial^\mu) A^\nu - \partial^\nu (\partial_\mu A^\mu)  = \mu_0 J^\nu$, which are explicitly Lorentz covariant and also gauge invariant, and the homogeneous Maxwell equations, given by the divergence of the dual tensor $\wt{F}^{\mu \nu} \equiv \vepsi^{\mu \nu \alpha \beta} F_{\alpha \beta}/2$, where $\vepsi^{\mu \nu \alpha \beta}$ is the permutation tensor, as $\partial_\mu \wt{F}^{\mu \nu} = 0$, are satisfied identically when written in terms of the electromagnetic potential, hence do not determine any degrees of freedom.  Antisymmetry in $F^{\mu \nu}$ immediately implies conservation of the current, $\partial_\nu \partial_\mu F^{\mu \nu} = \mu_0 \partial_\nu J^\nu = 0$, thus it carries only 3 degrees of freedom also.  One may recast the Maxwell equations into a component-free form through the use of differential geometry~[\cite{ryder-qft}], where ``the existence of integrals implies a duality between forms and chains'' which may be exploited.  In natural units $\mu_0 \equiv \vepsi_0 \equiv c \equiv 1$ and using the exterior derivative $\dext$, the Hodge dual $\;^*$, the connection 1-form $A \equiv A_\mu dx^\mu$, the curvature 2-form $F \equiv (-F_{\mu \nu}/2) dx^\mu \wedge dx^\nu$, and the current 3-form $J \equiv (J_x dy \wedge dz + J_y dz \wedge dx + J_z dx \wedge dy) \wedge dt - \rho_e dx \wedge dy \wedge dz$ which satisfies the continuity equation $\dext\, J = 0$, one writes the field equation as $\dext\, ^*\!F = J$ and the Bianchi identity, which is a statement on the structure of the manifold, as $\dext\, F = 0$, whence $F = \dext\, A$, and we remark that gauge invariance, through Noether's theorem~[\cite{noether-235,tavel-183}], implies conservation of the covariant current.  What all this shows is that the natural, physical division of the Maxwell equations is not into the divergence and curl equations but rather into the homogeneous and inhomogeneous equations, whereby the Bianchi identity carries the structure for the potential formulation and the field equation carries the dynamics obtained from the action.

The implication for plasma physics is clear: the quasineutral approximation does worse than just neglect an effect, as it introduces inconsistency into the equations when the components of the electrostatic field are treated in isolation~[\cite{rwj-ppcf04}].  Arguing that Maxwell's divergence equations are initial conditions for the curl equations is incorrect in media, for while in vacuum such statement leads to the propagation of electromagnetic radiation with two physical states of polarization, the source terms spoil such interpretation, and the divergence of the Maxwell-Ampere equation only recovers the equation for local charge conservation, which must be respected independently of the conservation of mass addressed by the zeroth moment of the Vlasov equation, when Gauss's law retains its intended form.  Note that authors including the no-monopole equation explicitly within the system do not make the argument of having 14 equations and degrees of freedom, as that equation represents an additional member.  Claiming that in general the sources may be uniquely determined from expressions for the fields is inappropriate, for while suitable boundary conditions must be supplied, the differential operators hence the boundary conditions are applied to the fields, not the sources.  The reason for the expression ``Maxwell-Lorentz electrodynamics'' is because Maxwell's theory tells one how the fields react to the sources, and the Lorentz force through the equations of motion tells the sources how to react to the fields; trying to go the other way around the loop is not well defined, as the physics is contained within the action from which both the field and source equations of motion may be obtained.  

Let us examine in detail where difficulties are encountered by the neoclassical approach, a term we use to encompass all non-classical approaches to the fluid description of ionized particles regardless of geometry---such discussion~[\cite{npg-12-425-2005,npg-14-49-2007}] invariably engenders a hostile response~[\cite{npg-14-543-2007,npg-14-545-2007}] from its adherents yet is necessary if one is to consider the application of electrodynamic field theory in tensor notation to the many-body system commonly called a plasma.  The scalar degrees of freedom $\rho_m$ and $p$ may be associated with the scalar equations for mass and energy conservation, Equations~(\ref{eqn:1}), as no other quantities appear in those equations for the case of vanishing flow velocity; the presence of a flow velocity $\mbf{V}_f$ couples those equations to the rest of the system to be solved simultaneously.  Note that the previous argument tacitly assumed that the equations of motion in the form of the generalized Ohm's law and the convective force balance, Equations~(\ref{eqn:2}), were associated with the degrees of freedom $\{\mbf{V}_f,\mbf{J}\}$; whereas here, without Gauss's law, one must determine the electric field from an equation of motion, usually the generalized Ohm's law (however the ion~[\cite{solomonetal-pop-2006}] and electron~[\cite{frc-pop-2006}] equations of motion are also used), giving the solution $\mbf{E}_{neo} = \eta \mbf{J} - \mbf{V}_f \times \mbf{B}$.    Faraday's law in conjunction with the no-monopole equation then relates the electric field to the potentials $-\mbf{E}_{neo} = \partial \mbf{A}/\partial t + \del \Phi$, where without Poisson's equation or its gauge invariant generalization the relation between the potentials and the space charge density $\rho_e$ remains unspecified (in essence, Faraday's law here {\it determines} a potential $\Phi$ which is not an independent degree of freedom), and its divergence gives in various gauges \bea
\divr \left( \mbf{V}_f \times \mbf{B} - \eta \mbf{J} \right) & = & \ddt{} \divr \mbf{A} + \del^2 \Phi \;, \\
\mathrm{Coloumb}\; (\divr \mbf{A} = 0)\; & = & \del^2 \Phi \;, \\
\mathrm{Lorenz}\; (\divr \mbf{A} = -\mu_0 \vepsi_0 \ddt{} \Phi)\; & = & \Box^2 \Phi \;, \\
\mathrm{Weyl}\; (\Phi = 0)\; & = & \ddt{} \divr \mbf{A} \;,
\eea where the LHS is explicitly gauge invariant whereas the form and interpretation of the RHS is dependent upon one's choice of gauge.  The issue of gauge invariance is a red herring in the discussion, for while true physics must be equally described in any and all gauges, the crucial error in the neoclassical approach is its use of an equation of motion to determine the electric field, which does not respect Lorentz covariance.  (Note that modern power generators and electric motors certainly are not moving materially at relativistic speeds yet make full and practical use of the covariant transformation properties of the field tensor through Faraday's law of induction.)  Returning to the expression for $\mbf{E}_{neo}$, let us now examine its transformation properties under a change of reference frame.  Let $S$ be the frame of the neoclassical observer, and let $S'$ be the frame moving with velocity $\mbf{V}_f$ with respect to $S$.  Without loss of generality, the flow velocity in $S$ is taken along the $x$-axis, thus $\mbf{V}_f= (V_f, 0, 0) \neq 0$ gives $\mbf{E}_{neo}=(\eta J_x, \eta J_y + V_f B_z, \eta J_z - V_f B_y)$, using Einstein's velocity addition rule~[\cite{einstein-1905a}] gives $\mbf{V}_f'=0$, and for $\gamma \equiv 1/\sqrt{1-V_f^2/c^2}$ the transformation for proper velocity applies to $\mbf{J}$, the spatial part of the four-current $J^\mu_{neo} = (0,\mbf{J})$, giving \beq
\mbf{E}_{neo}' = \eta' \mbf{J}' = \left[ \begin{array}{c} \eta' \gamma J_x \\ \eta' J_y \\ \eta' J_z \end{array} \right] \neq \left[ \begin{array}{c} \eta J_x \\ \gamma \eta J_y \\ \gamma \eta J_z \end{array} \right] = \mbf{E}' \;,
\eeq where $\mbf{E}' = [E_x, \gamma(E_y - V_f B_z), \gamma(E_z + V_f B_y)]$ is the transformation law for the physical electric field.  Equality could hold only if $\eta' = \eta / \gamma = \eta \gamma$ implying $\gamma = 1$, which holds only when $V_f = 0$, thus only in the neoclassical frame of reference but also implying a vanishing flow velocity, contradicting the initial assumption $V_f \neq 0$.  The expression for $\mbf{E}_{neo}$ has inherited the nature of a velocity vector from its neoclassical determination hence cannot possibly represent a true electric field, which {\it does not} transform as the spatial part of a four-vector~[\cite{maxwell-1864,griffiths-89,einstein-1905a}].  Furthermore, as ultimately $\mbf{B}(\mbf{J})$ may be determined from Ampere's law or the equivalent Biot-Savart law (for steady currents only as all the terms with $\mbf{E}$ need be present for Maxwell's theory to respect local charge conservation), the neoclassical electric field depends explicitly on the two vectorial quantities of current and mass flow, $\mbf{E}_{neo}(\mbf{J},\mbf{V}_f)$.  In order to completely determine the system, both of those quantities must find solution; however, having already used one of our equations of motion in the guise of Ohm's law, we have left only one vector equation for the net conservation of momentum, $\rho_m D \mbf{V}_f / D t + \del p = \mbf{J} \times \mbf{B}$, which leaves one vector's worth of degrees of freedom without solution, leading to the use of a stationary equilibrium equation $\del p = \mbf{J} \times \mbf{B}$ in the analysis of non-stationary plasma experiments~[\cite{frc-pop-2006,solomonetal-pop-2006}].  We note that the predictions of the neoclassical (NCLASS) model for the poloidal velocity found in a tokamak presented in Reference~[\cite{solomonetal-pop-2006}] explicitly fail to agree with the experimental measurements.  By reinstating the determination of the electrostatic field via Gauss's law, what returns is the generalized Ohm's law, an equation of motion which one may solve for the motion appearing in that equation, which in conjunction with the convective force balance equation fully determines the system.  Ultimately, the various arguments presented in support of the neglect of Gauss's law are superseded by the rigorous formalism of differential geometry, whereby casting the Maxwell equations into intrinsic, geometric form, $\dext\, ^* \dext\, A = J$, comprises very deep and powerful statements concerning what is known about our Universe.


\newpage

\begin{figure}
\includegraphics*[bb = 91 135 533 544, scale=.9]{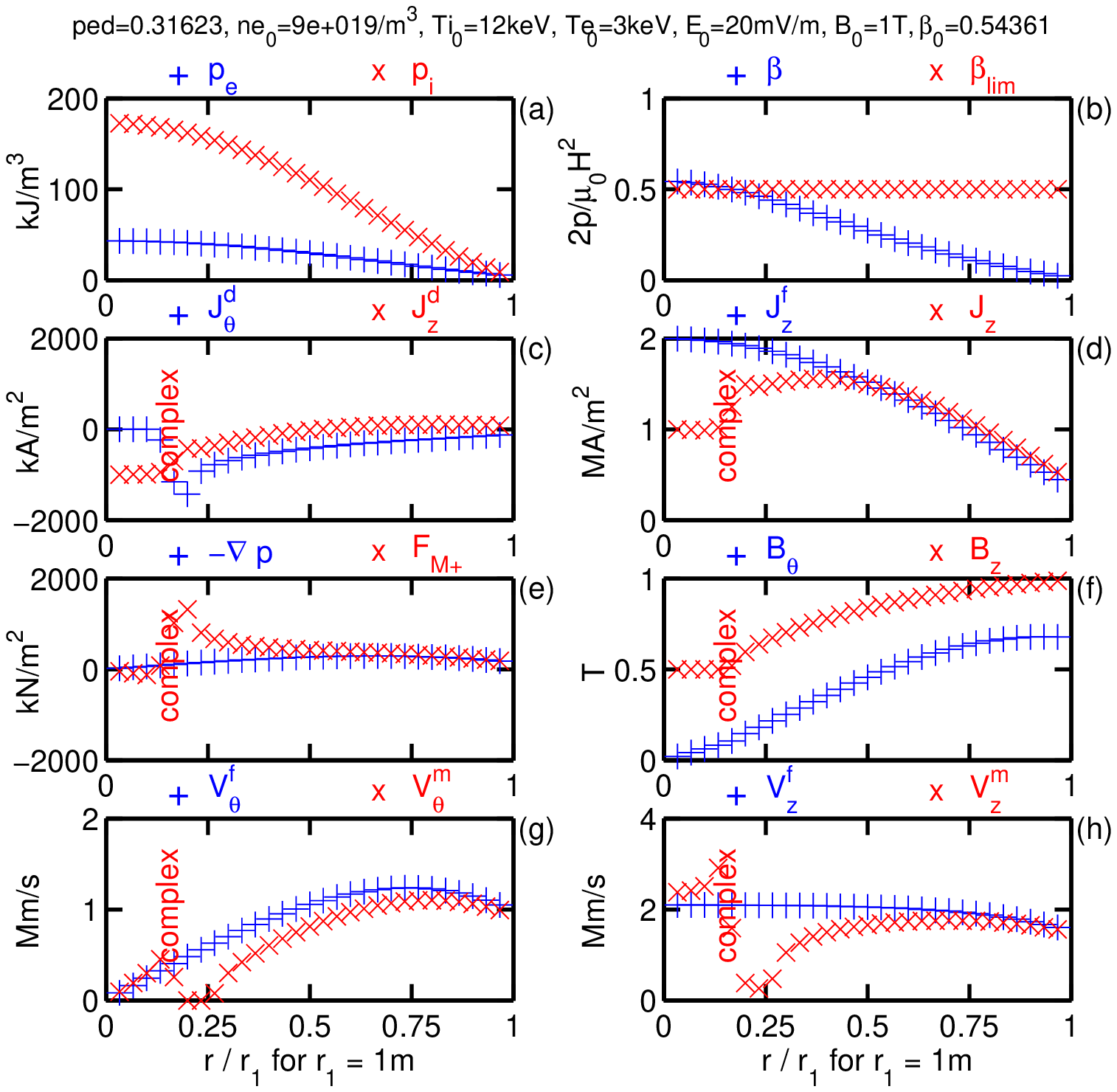}
\caption{\label{figF}(Color online.)  Solution profiles for $\chi_p = 1/\sqrt{10}$, $n_0 = 9 \times 10^{19}/{\rm m}^3$, $\Tbar_e = 3$keV, $\Tbar_i = 12$keV, $E_0 = 20$mV/m, $B_0 = 2$T, and $\beta_0 = 54.4\%$. (a) Electron pressure $+$ and ion pressure $\times$. (b) Kinetic to magnetic pressure ratio $\beta\;+$ and its limit $\beta_{\rm lim}\;\times$. (c) Diamagnetic current in azimuthal $+$ and axial $\times$ directions. (d) Axial free current $+$ and net current $\times$. (e) Gradient forces for pressure $- \del p \; +$ and magnetization $F_{M} \; \times$. (f) Net magnetic field in the azimuthal $+$ and axial $\times$ directions. (g) Azimuthal fluid velocity for the free current model $+$ and the magnetized model $\times$. (h) Axial fluid velocity for the free current model $+$ and the magnetized model $\times$.  Note that the magnetization has been driven complex near the plasma core, affecting the entire magnetization model while leaving the free current model unaffected.}
\end{figure}

\end{document}